\documentclass[11pt,a4paper]{article}
\usepackage{jheppub}
\pdfoutput=1

\usepackage{siunitx}
\usepackage{pdfpages}
\usepackage{float}
\usepackage{textcomp}               
\usepackage{braket,slashed,bm}
\usepackage{array,multirow}
\usepackage[normalem]{ulem}
\usepackage{xcolor,cancel,youngtab}

\usepackage[T1]{fontenc}
\usepackage{mathrsfs}
\usepackage{booktabs}
\usepackage{adjustbox}
\usepackage{mathtools}
\usepackage{hhline}
\usepackage{soul}

\usepackage{tikz}
\usetikzlibrary{arrows,decorations.pathmorphing,backgrounds,positioning,fit,petri,automata,shadows,calendar,mindmap,
decorations.markings,calc}

\definecolor{labelkey}{rgb}{0,0.5,0.0}

\usepackage{xspace}
\usepackage{slashed}
\usepackage{array,multirow}
\usepackage{graphicx}
\usepackage{graphics}
\usepackage{epsfig}
\usepackage{amsfonts}
\usepackage{amsmath}
\usepackage{amssymb}
\usepackage{dsfont}
\usepackage{color}
\usepackage{pdflscape}
\usepackage{pifont}
\usepackage{pgfplots}
\usepackage{tikz} 
\usepackage{pgfplotstable}
\usepackage{bbold}

\newcommand{\hc}{\mathrm{h.c.}}
\newcommand{\ep}{\epsilon}

\usepackage{bm}                                  

\newcommand{\beq}{\begin{equation}}
\newcommand{\eeq}{\end{equation}}
\newcommand{\be}{\begin{equation}}
\newcommand{\ee}{\end{equation}}
\newcommand{\bea}{\begin{eqnarray}}
\newcommand{\eea}{\end{eqnarray}}
\newcommand{\ben}{\begin{eqnarray*}}
\newcommand{\een}{\end{eqnarray*}}

\newcommand{\simle}{\hspace*{0.2em}\raisebox{0.5ex}{$<$}
     \hspace{-0.8em}\raisebox{-0.3em}{$\sim$}\hspace*{0.2em}}

\newcommand{\bma}{\begin{pmatrix}}
\newcommand{\ema}{\end{pmatrix}}

\def\lixo#1{}
	
\def\simle{\hspace*{0.2em}\raisebox{0.5ex}{$<$}
     \hspace{-0.8em}\raisebox{-0.3em}{$\sim$}\hspace*{0.2em}}

\def\slashchar#1{\setbox0=\hbox{$#1$}           
  \dimen0=\wd0                                    
  \setbox1=\hbox{/} \dimen1=\wd1                  
  \ifdim\dimen0>\dimen1                           
    \rlap{\hbox to \dimen0{\hfil/\hfil}}            
    #1                                             
  \else                                          
    \rlap{\hbox to \dimen1{\hfil$#1$\hfil}}        
    /                                           
 \fi}                                           %

\newcommand{\dslash}[1]{#1 \llap{/\kern-0.5pt}}
\newcommand{\Dslash}[1]{#1 \llap{/\kern+1.5pt}}
\newcommand{\DDslash}[1]{#1 \llap{/\kern+2.3pt}}
\newcommand{\dslashh}[1]{#1 \llap{/\kern+1pt}}

\newcommand{\nn}{\nonumber}

	\preprint{\begin{flushright} BONN-TH-2021-11
	\end{flushright}}	

	\title{Long-lived Sterile Neutrinos at Belle II in Effective Field Theory}

		\author[a,b]{Guanghui~Zhou,}
		\emailAdd{g.zhou@uva.nl}
		\affiliation[a]{Institute for Theoretical Physics Amsterdam and Delta Institute for Theoretical Physics, University of Amsterdam, Science Park 904, 1098 XH Amsterdam, The Netherlands}
		\affiliation[b]{Nikhef, Theory Group, Science Park 105, 1098 XG, Amsterdam, The Netherlands}
				
		\author[c]{Julian~Y.~G\"unther,}
		\emailAdd{guenther@physik.uni-bonn.de}
		\affiliation[c]{Bethe Center for Theoretical Physics \& Physikalisches Institut der Universit\"at Bonn,\\ Nu{\ss}allee 12, 53115 Bonn, Germany}

		\author[d,e]{Zeren~Simon~Wang,}
		\emailAdd{wzs@mx.nthu.edu.tw}
		\affiliation[d]{Department of Physics, National Tsing Hua University, Hsinchu 300, Taiwan}
		\affiliation[e]{Center for Theory and Computation, National Tsing Hua University, Hsinchu 300, Taiwan}
		
		\author[a,b]{Jordy~de~Vries,}
		\emailAdd{j.devries4@uva.nl}

		\author[c]{and Herbi~K.~Dreiner}
		\emailAdd{dreiner@uni-bonn.de}

\abstract{Large numbers of $\tau$ leptons are produced at \texttt{Belle II}. These could potentially decay into sterile neutrinos that, for the 
mass range under consideration, are typically long-lived, leading to displaced-vertex signatures. Here, we study a displaced-vertex search in the context of the
sterile-neutrino-extended Standard Model Effective Field Theory. 
The production and decay of the sterile neutrinos can be realized via either the standard active-sterile neutrino mixing or higher-dimensional 
operators in the effective Lagrangian. We perform Monte-Carlo simulations to estimate the \texttt{Belle II} sensitivities to such interactions.
We find that \texttt{Belle II} can probe non-renormalizable dimension-six operators involving a single sterile neutrino up to a few TeV in the 
new-physics scale.}

\begin{document}
\maketitle
			
\section{Introduction}\label{sec:introduction}
\setcounter{footnote}{0} 

The observation of neutrino oscillations has established active neutrinos as massive particles~\cite{deSalas:2020pgw}, inconsistent with the 
prediction of the Standard Model (SM) of particle physics. The $SU(2)_L \times U(1)_Y$ gauge invariance and the absence of  right-handed 
neutrinos forbid a renormalizable neutrino mass term. Thus, new physics (NP) is required to explain the neutrino masses. A minimal solution 
is to add right-handed gauge-singlet sterile neutrinos to the 
SM~\cite{Minkowski:1977sc,Yanagida:1979as,Mohapatra:1979ia,GellMann:1980vs,Schechter:1980gr}, which can interact with the SM particles 
via Yukawa interactions, generating Dirac neutrino masses after electroweak (EW) symmetry breaking. Without violating Lorentz symmetry or 
SM gauge symmetries, the sterile neutrinos can also have a Majorana mass term, which violates lepton number by two units and can result in 
lepton-number-violating processes such as neutrinoless double beta decay \cite{Dekens:2020ttz}. With a so-called seesaw mechanism, tiny 
active neutrino masses, in agreement with observations~\cite{deSalas:2020pgw,Esteban:2020cvm,Capozzi:2017ipn}, arise from $\mathcal{O}(1)$ Yukawa couplings and sterile neutrino masses of 
GUT (Grand Unified Theory) scale $\sim10^{15}\,$GeV. But much smaller sterile masses are equally fine, just requiring smaller Yukawa couplings. 

Besides being a minimal solution, sterile neutrinos are also predicted in a wide range of beyond-the-Standard-Model (BSM) models, such as
GUTs~\cite{Bando:1998ww}, $Z^\prime$ models \cite{Deppisch:2019kvs,Chiang:2019ajm}, left-right symmetric models 
\cite{Mohapatra:1974gc,Pati:1974yy,Mohapatra:1980yp,Keung:1983uu}, and leptoquark models \cite{Dorsner:2016wpm}. Beyond the sterile 
neutrinos, these models predict new particles that are often heavy compared to the EW scale. One useful and systematic way to describe such 
new physics is to apply effective field theory (EFT), where the heavy degrees of freedom are integrated out, leading to non-renormalizable 
operators in the Lagrangian. These are gauge invariant and consist of light fields only. The EFT that describes the interaction of the sterile 
neutrinos with SM particles is known as 
$\nu$SMEFT~\cite{delAguila:2008ir,Aparici:2009fh,Liao:2016qyd,Bell:2005kz,Graesser:2007yj,Graesser:2007pc}, which we apply in this work.

If the sterile neutrinos are light, \textit{e.g.}~at the GeV scale, and their mixings with the active neutrinos are tiny and/or the NP scale is 
heavy, their decay rates are suppressed and the sterile neutrinos become long-lived. If such a neutrino is produced at an experiment it can travel a 
macroscopic distance, before decaying into, hopefully, visible final states. Searches for such sterile neutrinos have been both performed and 
proposed at various experimental facilities. Past experiments including \texttt{Belle}~\cite{Belle:2013ytx}, 
\texttt{PS191}~\cite{Bernardi:1987ek,Ruchayskiy:2011aa}, \texttt{L3}~\cite{L3:2001zfe}, \texttt{T2K}~\cite{T2K:2019jwa}, 
\texttt{CHARM}~\cite{CHARM:1985nku,Orloff:2002de}, \texttt{NuTeV}~\cite{NuTeV:1999kej}, \texttt{NA3}~\cite{NA3:1986ahv}, 
\texttt{BEBC}~\cite{WA66:1985mfx}, and \texttt{DELPHI}~\cite{DELPHI:1996qcc} have attained constraints for sterile neutrino masses below 
the $W$-boson mass. More recently, \texttt{LHCb}~\cite{LHCb:2016inz,Antusch:2017hhu}, and \texttt{CMS}~\cite{CMS:2015qur,CMS:2018iaf} 
have also searched for such exotics. In addition, \texttt{ATLAS} at the LHC has searched for sterile neutrinos which mix with either $\nu_e$ or 
$\nu_\mu$~\cite{ATLAS:2015gtp,ATLAS:2019kpx}. Furthermore, a number of far-detector experiments such as 
\texttt{FASER}~\cite{FASER:2018eoc} and \texttt{MATHUSLA}~\cite{Curtin:2018mvb} have been proposed to be operated in the vicinity of 
various LHC interaction points (IPs). These are planned as dedicated detectors to hunt for long-lived particles (LLPs) in 
general\footnote{See Refs.~\cite{Curtin:2018mvb,Lee:2018pag,Alimena:2019zri,Beacham:2019nyx} for recent reviews of LLP models and searches.}, and their 
sensitivities to long-lived sterile neutrinos have been studied extensively~\cite{Kling:2018wct,Hirsch:2020klk,Cottin:2021lzz,Helo:2018qej,Dercks:2018wum,DeVries:2020jbs,Deppisch:2019kvs}. Besides, excellent 
sensitivities to such scenarios are also expected at future electron-positron and electron-proton colliders such as the CEPC, 
FCC-ee, LHeC, and FCC-he~\cite{Das:2018usr,Cottin:2021tfo,Antusch:2019eiz,Fischer:2017wkj,Antusch:2016ejd,Antusch:2016vyf,Antusch:2015mia}.
Finally, $B$-factories such as the ongoing \texttt{Belle~II} experiment, colliding electron and positron beams at relatively low center-of-mass (CM) 
energies, could also look for sterile neutrinos lighter than $B$-mesons~\cite{Belle-II:2018jsg,Belle:2013ytx,Kobach:2014hea,Dib:2019tuj}.

In this work, we focus on the $B$-factory experiment \texttt{Belle~II}, which is in operation in Japan. At \texttt{Belle~II}, an electron beam of energy 7\,GeV collides with a positron beam of energy 4\,GeV, reaching the CM energy 10.58\,GeV, \textit{i.e.}~at the $\Upsilon(4S)$ resonance. With a projected integrated luminosity of 50 ab$^{-1}$, this results in a very large number of $B\bar{B}$ events. Besides, \texttt{Belle~II} is estimated to 
generate a large number of $\tau$-pair production events, allowing for the study of rare $\tau$ decays to an unprecedented precision. This includes 
studying lepton flavor violation~\cite{Heeck:2017xmg,Tenchini:2020njf,Cheung:2021mol,Daub:2012mu,Dreiner:2012mx} and LLPs~\cite{Kim:2019xqj,Dib:2019tuj,Duerr:2019dmv,Dey:2020juy,Duerr:2020muu,Filimonova:2019tuy,Chen:2020bok,Cheung:2021mol,Bertholet:2021hjl,Kang:2021oes,Acevedo:2021wiq,Dreyer:2021aqd}.
In particular, Ref.~\cite{Dib:2019tuj} has studied the \texttt{Belle~II} exclusion limits for long-lived sterile neutrinos which mix dominantly with $\nu_\tau$,
by considering $\tau$ decays. Here, we propose a displaced-vertex search strategy similar to that discussed in Ref.~\cite{Dib:2019tuj}, reproducing the minimal-scenario results, as well as extending the physics coverage to several scenarios in the $\nu$SMEFT, where one single EFT operator can lead to both production and decay of the sterile neutrinos.

The paper is organized as follows. We first introduce the $\nu$SMEFT theoretical framework in Sec.~\ref{sec:model}. We discuss both the minimal scenario and a series of EFT scenarios in Sec.~\ref{sec:scenarios}, for all of which we  perform numerical studies. The \texttt{Belle~II} experiment is 
detailed in Sec.~\ref{sec:simulation} together with a description of our search strategy. The numerical results are presented in Sec.~\ref{sec:results}, and we conclude the paper in Sec.~\ref{sec:conclusions}.
Appendices~\ref{appendix:twobodydecay} and~\ref{appendix:threebodydecay} detail the computation of two- and three-body decay rates of the sterile neutrinos.

\section{The $\nu$SMEFT Model}\label{sec:model}

For simplicity, we consider the SM extended by only one right-handed gauge-singlet neutrino $\nu_R$.
The SM Lagrangian is then augmented by new renormalizable
terms: a Majorana mass term and a Yukawa term
\begin{eqnarray}\label{smeftdim4}
	\mathscr L &=&  \mathscr L_{\text{SM}} - \left[ \frac{1}{2} \bar \nu^c_{R} \,\overline{M}_R \nu_{R} +\bar L \tilde H Y_\nu \nu_{R} + \rm{h.c.}\right]\,,
\end{eqnarray}
where $\mathscr L_{\text{SM}}$ denotes the SM Lagrangian, $L=(\nu_{L}, e_{L})^T$ is the lepton doublet, $H$ is the Higgs doublet, and $\tilde H = i \tau_2 H^*$.
In the unitary gauge, 
\begin{equation}
	H = \frac{v}{\sqrt{2}}\rm  \left(\begin{array}{c}
		0 \\
		1 + \frac{h}{v}
	\end{array} \right)\,,
\end{equation}
where $v$ = 246\,GeV is the Higgs vacuum expectation value and $h$ is the SM Higgs scalar. $\overline{M}_R$ is the Majorana mass for the 
sterile neutrino and $Y_\nu$ is a 3$ \times 1$ matrix of Yukawa couplings. We work in the basis where the charged leptons $e^i_{L,R}$, the 
quarks $u^i_{L,R}$, and $d^i_R$ ( $i=1,2,3$) are in their mass eigenstates. For $d^i_L$ we have $d^i_L = V^{ij} d_L^{j,\,\rm mass}$, where 
$V$ is the CKM matrix and $d_L^{j,\,\rm mass}$ denotes the left-handed down-type quarks in the mass basis. $\nu^c_R$ is the charge 
conjugate field of $\nu_R$ with $\nu^c_R = C\bar{\nu}^T_R$ and $C = -i\gamma^2\gamma^0$, in four-component fermion notation.

Additional new physics at a higher energy scale can lead to higher-dimensional operators involving the $\nu_{R}$ above the electroweak scale. The possible dimension-5 operators are 
\be
\mathcal L^{(5)}_{\nu_L} =  \ep_{kl}\ep_{mn}(L_k^T\, C_L^{( 5)}\,CL_m )H_l H_n\,,\qquad  
\mathcal L^{(5)}_{\nu_R}=- \bar \nu^c_{R} \,C_R^{(5)} \nu_{R} H^\dagger H\,.
\label{eq:dim-5-operators}
\ee
Here, $C$ is the charge conjugation matrix, as before. $C_L^{(5)}$ and $\,C_R^{(5)}$ are arbitrary coefficients with mass dimension -1. The first 
term is also known as the Weinberg operator. Both terms contribute to the Majorana masses for the active and sterile neutrinos, respectively, after 
electroweak symmetry breaking (EWSB). Thus, at low energy, these terms only lead to a shift in the free parameter $\overline M_R$, and are not 
relevant for our analysis.

Here, we are interested in operators with just one sterile neutrino up to dimension six~\cite{Liao:2016qyd,Grzadkowski:2010es}. We list
the relevant ones in Table~\ref{tab:O6R}\footnote{In principle, there are also dimension-six operators with two or four sterile neutrinos. See 
Ref.~\cite{Liao:2016qyd} for a summary. In particular, operators with a pair of sterile neutrinos can enhance the production of the sterile 
neutrinos, which then may dominantly decay via mixing with the SM active neutrinos. See Ref.~\cite{Cottin:2021lzz} for a recent study on 
this scenario at the LHC. 
}. They are organized in four classes according to powers of the fermion 
($\psi$) and Higgs fields (H): $\psi^2H^3,\,\psi^2H^2D,\psi^2HF,\,\psi^4$, where $D$ denotes a covariant derivative and $F$ is a gauge field 
strength tensor. We use $C_{I}$ to denote the Wilson coefficients of operator $\mathcal O_I$. Furthermore, each carries two or four generation
indices for the quarks and leptons. 
{\renewcommand{\arraystretch}{1.3}\begin{table}[t!]\small
		\center
		\begin{tabular}{||c|c||c|c||}
			\hhline{|====|} \textbf{Class 1}& $\psi^2 H^3$  & \textbf{Class 4} &  $\psi^4 $\\
			\hline
			$\mathcal{O}^{}_{L\nu H}$ & $(\bar{L}\nu_R)\tilde{H}(H^\dagger H)$ & $\mathcal{O}^{}_{du\nu e}$ & $ (\bar{d}_R
			\gamma^\mu u_R)(\bar{\nu}_R \gamma_\mu e)$  
			\\ \hhline{||==||~|~||} 
			\textbf{Class 2}&  $\psi^2 H^2 D$ &  $\mathcal{O}^{}_{Qu\nu L}$ & $(\bar{Q}u_R)(\bar{\nu}_RL)$  \\ \cline{1-2}  
			$\mathcal{O}^{}_{H\nu e}$ & $(\bar{\nu }_R\gamma^\mu e_R)({\tilde{H}}^\dagger i D_\mu H)$ & $\mathcal{O}^{}_{L\nu Qd}$ 
			& $(\bar{L}\nu_R )\epsilon(\bar{Q}d_R)$
			\\ \hhline{||==||~|~||}
			\textbf{Class 3} &  $\psi^2 H F$  & $\mathcal{O}^{}_{LdQ\nu }$ & $(\bar{L}d_R)\epsilon(\bar{Q}\nu_R )$ \\ \cline{1-2}
			$\mathcal{O}^{}_{\nu W}$ &$(\bar{L}\sigma_{\mu\nu}\nu_R )\tau^I\tilde{H}W^{I\mu\nu}$  & $\mathcal{O}^{}_{L\nu Le}$ 
			&$(\bar{L}\nu_R)\epsilon (\bar{L}e_R)$ \\ \cline{1-2}
			$\mathcal{O}^{}_{\nu B}$ &$(\bar{L}\sigma_{\mu\nu}\nu_R )\tilde{H}B^{\mu\nu}$& & \\
			\hhline{|====|}
		\end{tabular}
		\caption{Dimension-six operators involving one sterile neutrino field $\nu_R$.
		Each fermion field has a generation index, except $\nu_R$.
		When needed, we shall attach these indices to the operator symbol.
		Thus, $\mathcal{O}^{21\nu_R3}_{du\nu e}$ refers to the operator $(\bar{s}\gamma^\mu u)(\bar{\nu}_R \gamma_\mu \tau)$.
		} \label{tab:O6R}
\end{table}}

After EWSB, the operator $\mathcal O^{}_{L\nu H}$ contributes to the Dirac mass of the neutrino. This can be absorbed in a re-definition of $Y_\nu$ 
and will therefore not be considered for  the rest of the paper. $\mathcal O_{\nu W}$ and $\mathcal O_{\nu B}$ induce higher-dimensional operators 
with much smaller coefficients compared  to the other operators~\cite{Dekens:2020ttz,DeVries:2020jbs} listed in Table~\ref{tab:O6R} and are strictly 
constrained by neutrino dipole moments~\cite{Butterworth:2019iff,Canas:2015yoa}, so will not be considered further. The remaining interactions are 
gauge invariant under $SU(3)_C \times U(1)_{\text{EM}}$ and can be written as 
\begin{eqnarray}\label{d6total}
	\mathscr L_{} &=&  \mathcal L_{\text{SM}}-  \left[\frac{1}{2} \bar \nu^c_{L} \, M_L \nu_{L}   +  \frac{1}{2} \bar \nu^c_{R} \, M_R \nu_{R} +\bar \nu_L M_D\nu_R +\hc \right]\nn \\
	&&+\mathcal L^{(6)}_{CC} +\mathcal L^{(6)}_{NC}\,,
	\label{eq:eff-lagrangian-all}
\end{eqnarray}
where $\mathcal L_{\text{SM}}$ denotes the renormalizable Lagrangian involving only light SM fields after EWSB.
$M_L$ is a $3\times3$ Majorana mass matrix, $M_R$ is a Majorana mass parameter, and $M_D$ is a $3\times1$ Dirac mass matrix.
$\mathcal L^{(6)}_{CC}$ contains charged-current interactions and is given by 
\bea\label{d6CC}
\mathcal L^{(6)}_{CC}& =& \frac{2 G_F}{\sqrt{2}} \Bigg\{ 
\bar u_L^{i} \gamma^\mu d_L^{j} \left[  \bar e_{L}^{k}   \gamma_\mu c^{CC}_{\textrm{VLL}, ijkl} \,  \nu_{L}^{l} + \bar e_{R}^{k}   
\gamma_\mu  c^{CC1}_{\textrm{VLR},ijk} \,  \nu_{R} \right]+
\bar u_R^{i}  \gamma^\mu d_R^{j} \,\bar e_{R}^{k} \,  \gamma_\mu   c^{CC}_{\textrm{VRR}, ijk} \,\nu_{R}\nn \\
& & +
\bar u_L^{i}   d_R^{j}  \,\bar e_{L}^{k} \,  c^{CC1}_{ \textrm{SRR}, ijk}  \nu_{R} + 
\bar u_R^{i}   d_L^{j}  \, \bar e_{L}^{k}  \,  c^{CC}_{ \textrm{SLR}, ijk}    \nu_{R} +  \bar u_L^{i}  \sigma^{\mu\nu} d_R^{j} \,  
\bar e_{L}^{k}   \sigma_{\mu\nu}  c^{CC}_{ \textrm{T}, ijk} \, \nu_{R}\nn\\
& & +  \bar{e}_L^{i}  c^{CC2}_{\textrm{SRR}, ijk}\nu_R \bar{\nu}_L^{j}  e^{k}_{R}   +\bar{\nu}^i_L\gamma^\mu e^j_L \bar{e}_R^{k} \gamma_\mu c^{CC2}_{\textrm{VLR}, ijk}\nu_R  \Bigg\}  +{\rm h.c.}
- \frac{4G_F}{\sqrt{2}}\bar{\nu}^i_L\gamma^\mu e^i_L \bar{e}^j_L \gamma_\mu \nu^j_L \label{lowenergy6_l0},
\eea
where we include terms involving only active neutrinos $\nu_L$ from the SM weak interaction and $i,j, k, l$ are the flavor indices of the quarks and leptons, 
and a summation over them is implied. 
Similarly, for the neutral-current interactions $\mathcal L^{(6)}_{NC}$, we have 
\begin{equation}\label{d6NC}
	\begin{aligned}
		\mathcal L^{(6)}_{\rm NC}&=\frac{-4 G_F}{\sqrt{2}}\bar \nu^i_L \gamma^\mu \nu^i_L\Bigg\{ 
		\bar e^j_{L}  \gamma_\mu (-\frac{1}{2}+\sin^2\theta_W)  e^j_L+ \bar e^j_{R}  \gamma_\mu (\sin^2\theta_W) e^j_R \\
		& +  \bar u^j_L \gamma^\mu (\frac{1}{2}-\frac{2}{3}\sin^2\theta_W)u^j_L\,+\bar u^j_{R}\,  \gamma_\mu   (-\frac{2}{3}\sin^2\theta_W)u^j_R \\
		& +  \bar d^j_L \gamma^\mu (-\frac{1}{2}+\frac{1}{3}\sin^2\theta_W)d^j_L\,+\bar d^j_{R}\,  \gamma_\mu   (\frac{1}{3}\sin^2\theta_W)d^j_R 
		+ \frac{1}{4}(2-\delta_{ij})\bar \nu^j_L \gamma^\mu \nu^j_L \Bigg\}\, \\
		&+\frac{2G_F}{\sqrt{2}}\Bigg\{ \bar{u}_R^{i}  u_L^{j} \bar{\nu}_L^{k}  c^{NC}_{\textrm{SLR}, ijk}\nu_R+ 
		\bar{d}_L^{i} d_R^{j} \bar{\nu}_L^{k}  c^{NC1}_{\textrm{SRR},ijk}\nu_R +\bar{\nu}_L^{i}  c^{NC2}_{\textrm{ SRR}, ijk}\nu_R\bar{e}_L^{j} e_R^{k}  \\
		&+\bar{d}_L^{i} \sigma^{\mu\nu}d_R^{j}  \bar{\nu}_L^{k} \sigma_{\mu\nu} c^{NC}_{\textrm{T}, ijk}\nu_R +\rm h.c. \Bigg\}, 
	\end{aligned}
\end{equation}
where $\theta_W$ is the electroweak mixing angle.
Here, we give the matching relations~\cite{Dekens:2020ttz}.
For the mass terms  we find 
\bea\label{Massmatch}
M_L = -v^2 C_L^{(5)}\,,\qquad
M_R = \overline M_R + v^2  C_R^{(5)}\,,\qquad
M_D =\frac{v}{\sqrt{2}} \left[Y_\nu -\frac{v^2}{2}C_{L\nu H}^{}\right]\,.
\eea 
 The matching relations for  the charged-current  operators are
\bea\label{match6CC}
c_{\textrm{VLL}, ijkl}^{CC} &=& -2V_{ij} \delta_{kl}\,,\qquad \qquad \qquad \qquad \qquad\;\;
c_{\textrm{VLR}, ijk}^{CC1} = \left[-v^2C_{H\nu e, k}^{}\right]^\dagger V_{ij}\,,\nn\\
c_{\textrm{VLR}, ijk}^{CC2} &=& \left[-v^2C_{H\nu e, k}\right]^\dagger \delta_{ij}\,, \qquad \qquad \quad \quad\;\;\,
c_{\textrm{ VRR}, ijk}^{CC} = v^2\left(C_{du\nu e, jik}^{}\right)^\dagger\,,\nn\\
c_{\textrm{SRR},ijk}^{CC1}&=& -v^2C_{L\nu Qd, kij}^{}+\frac{v^2}{2} C_{LdQ\nu,kji }^{}\,,\qquad
c_{\textrm{ SLR}, ijk}^{CC}= v^2\left(C_{Qu\nu L, lik}^{}\right)^\dagger V_{lj}\,,\nn\\
c_{\textrm{ T}, ijk}^{CC} &=& \frac{v^2}{8} C_{LdQ\nu,kji }^{}\,, \qquad \qquad\qquad \quad \qquad
c_{\textrm{ SRR},ijk}^{CC2}= -v^2 C_{L\nu Le,ijk}\,,
\eea
with $V$  the CKM matrix. 
For the neutral-current operators we find 
\bea\label{match6NC}
c_{\textrm{ SLR}, ijk}^{NC} &=& v^2\left(C_{Qu\nu L,jik}^{}\right)^\dagger \,, \qquad 
c_{\textrm{ SRR},ijk}^{NC1} =  v^2C_{L\nu Qd,k lj}^{}V_{li}^*-\frac{v^2}{2} C_{LdQ\nu,kjl }V_{li}^*\,,\nn\\
c_{\textrm{ SRR}, ijk}^{NC2} &=& v^2 C_{L\nu Le,ijk}^{}\,,\qquad \qquad \;
c_{\textrm{ T},ijk}^{NC} = -\frac{v^2}{8} C_{LdQ\nu,kjl }^{} V_{li}^*\,,
\eea
where $V_{li}^*$ is the charge conjugate of $V_{li}$.
 
The renormalization group equations for these $\nu$SMEFT operators arising from one-loop QCD effects have been discussed in Ref.~\cite{Dekens:2020ttz}. The overall effect there was found to be minor and we neglect the effect here.

\section{Theoretical Scenarios}\label{sec:scenarios}
In this section, we study various possible channels for the production and decay of sterile neutrinos at 
\texttt{Belle~II}. We extend the SM by adding a Majorana sterile neutrino $\nu_R$, as well as the non-renormalizable interactions given in Table~\ref{tab:O6R}. 
The sterile neutrino  is produced by the decay of $\tau$ leptons either from the mixing between 
active and sterile neutrinos, or from the new higher-dimensional operators. 

\subsection{The minimal Scenario}\label{subsec:minimal}
We first consider the minimal model, where a sterile neutrino interacts with SM particles only through the 
mixing  between the active tau-neutrino $\nu_\tau$ and the sterile neutrino. The Lagrangian is obtained by 
setting all the Wilson coefficients $c^{CC}$ and $c^{NC}$ except $c^{CC}_{\rm VLL}$ in Eqs.~\eqref{d6CC} 
and \eqref{d6NC} to zero. The active neutrinos $\nu_\alpha$ can be expressed in terms of neutrino mass 
eigenstates $\nu_i$
\begin{equation}
\nu_\alpha = U_{\alpha i}\nu_i\,,
\end{equation}
where $\alpha= e, \mu, \tau$ and $i= 1, 2, 3, 4$.
We only consider the mixing between $\nu_\tau$ and the sterile neutrino proportional to $U_{\tau 4}$ as the 
mixing matrix elements $U_{e4}$ and $U_{\mu4}$ are severely constrained from other observables, see \textit{e.g.} Refs.~\cite{Bondarenko:2018ptm, DeVries:2020jbs}. We neglect the masses of the active neutrinos. As a 
result of the small mixing $U_{\tau 4}$, $\nu_R$ is approximately equivalent to $\nu_4$, and we 
hence refer to both of them as the sterile neutrino. The production and decay rates depend on two 
independent parameters, the mixing matrix element $U_{\tau 4}$ and the mass of the sterile neutrino $m_N$. We list 
all the possible resulting production and decay channels in the minimal scenario in the second row of Table~\ref{tab:modes}. For instance, with $X_1=K^-$ and $X_2=e^-+e^+ +\nu_
\tau$, we have the decay chain,
\begin{equation}
\tau^- \rightarrow \nu_R+K^-,\qquad \mathrm{followed\;by,}\qquad \nu_R\rightarrow e^-+e^+ +\nu_\tau\,.
\end{equation}

{\renewcommand{\arraystretch}{1.3}\begin{table}[t!]\small
		\center
		\begin{tabular}{|c||c||c|}
			
			\hline  $\tau^- \rightarrow \nu_R\,  (\bar{\nu}_R)+ X_1$ $\&$ $\nu_R\rightarrow X_2  $ & $ X_1$  &      $ X_2  $\\
			\hline
			 Minimal scenario & $\pi^-, \rho^-, K^-, K^{*-},$  &  $(\pi^0, \rho^0, \eta, \eta^\prime, \omega, \phi,$  \\ 
			&  $e^- + \bar{\nu}_e\,, \mu^-+\bar{\nu}_\mu$ &    
			$ \bar{\nu}_e+\nu_e\,, \bar{\nu}_\mu+\nu_\mu\,, \bar{\nu}_\tau+\nu_\tau,$\\
			&  &$e^-+ e^+\,,\mu^-+\mu^+) + \nu_\tau$
			\\	\hline
		Scenario $\mathcal O^{3\nu_R11}_{ L\nu Qd} $ & $\pi^-$ 	& $(\pi^0\,, \eta\,, \eta^\prime\,, K^0) + \nu_\tau$   \\ \hline
		Scenario $\mathcal O^{11\nu_R3}_{Qu\nu L} $ & $\pi^-\,, K^-$ &	  $(\pi^0\,, \eta\,, \eta^\prime) + \nu_\tau$   \\ \hline
		 Scenario $\mathcal O^{\nu_R1}_{H\nu e}$ &  $e^- + \nu_\tau \, (+ \bar{\nu}_R)$ & ( $\pi^+, \rho^+, K^+, K^{*+},$   \\ 
			&   &$e^+ + \nu_e\,, \mu^++\nu_\mu) + e^-$ 
			\\	\hline
			
		 Scenario $\mathcal O^{11\nu_R3}_{du\nu e}$	$\&$ $\mathcal O^{11\nu_R1}_{du\nu e}$	& $\pi^-\,, \rho^-$ &   $(\pi^+\,, \rho^+)+e^-$
			\\ \hline
		 Scenario $\mathcal O^{1\nu_R31}_{ L\nu Le} $ & $e^- +\bar{\nu}_e$& 
			 $e^- + \nu_\tau +e ^+$ \\ \hline
		 Scenario $\mathcal O^{311\nu_R}_{ LdQ\nu} $ & $\pi^-\,, \rho^-$  & $(\pi^0\,, \rho^0\,, \omega\,, \eta\,, \eta^\prime\,, K^0\,, K^{*0})+\nu_\tau$\\ \hline

		\end{tabular}
		\caption{All possible production, $X_1$, and decay, $X_2$, modes of a sterile neutrino $\nu_R$ at \texttt{Belle~II}. 
			The charge conjugate modes are implied.
		} \label{tab:modes}
\end{table}}

\begin{figure}[t]
	\centering
	\includegraphics[width=0.8\textwidth]{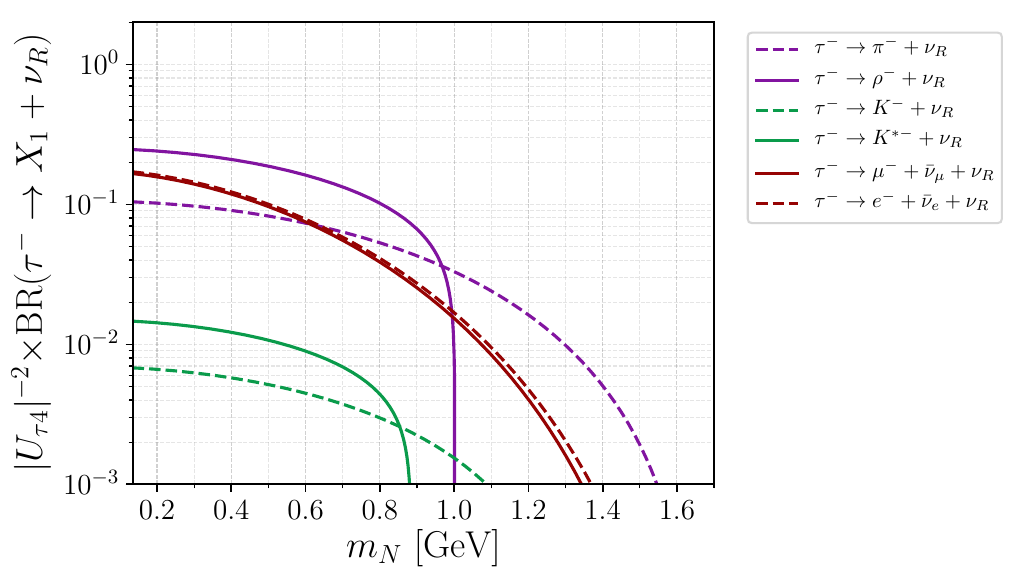}
	\caption{Tau decay branching ratios into a sterile neutrino in the minimal scenario.} \label{fig:BRSMtau}
\end{figure}

\begin{figure}[t]
	\centering
	\includegraphics[width=0.8\textwidth]{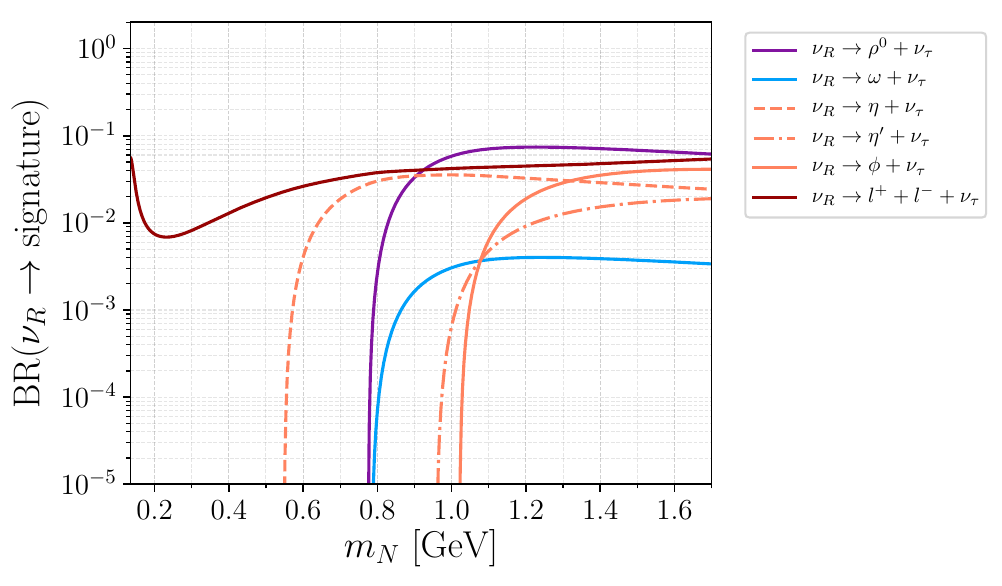}
	\caption{Branching ratios of visible decay modes in the minimal scenario.} \label{fig:SMBRN}
\end{figure}

\begin{figure}[t]
	\centering
	\includegraphics[width=0.6\textwidth]{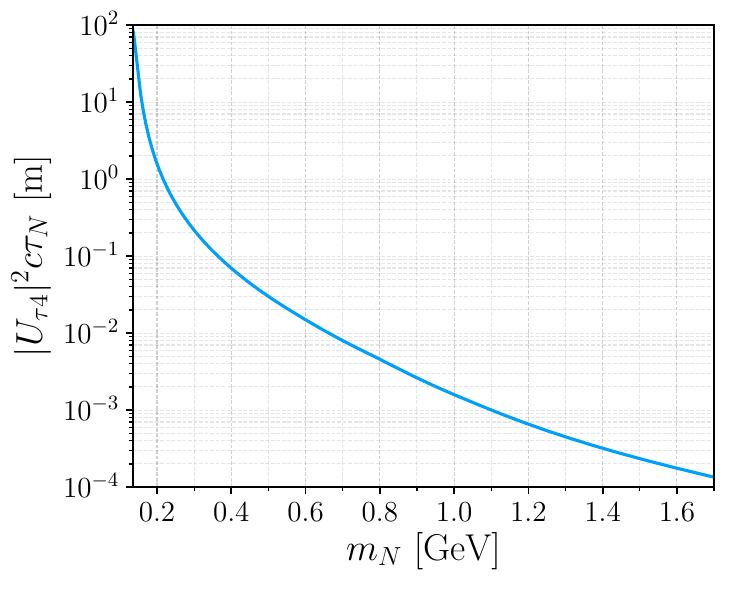}
	\caption{Proper decay length of the sterile neutrino times $|U_{\tau 4}|^2$ in the minimal scenario.} \label{fig:SMctau}
\end{figure}

The branching ratios of the production and decay channels are displayed in Fig.~\ref{fig:BRSMtau} and Fig.~\ref{fig:SMBRN}, as functions of $m_N$.
	As explained in Sec.~\ref{sec:simulation}, we do not consider all the decay channels of $\nu_R$ as visible.
	Therefore, in Fig.~\ref{fig:SMBRN} only the visible channels are shown.
	The resulting proper decay length of the sterile neutrino times $|U_{\tau 4}|^2$ is presented in Fig.~\ref{fig:SMctau}.
	For a discussion of the explicit expression of the decay rates, we refer to the Appendices~\ref{appendix:twobodydecay} and \ref{appendix:threebodydecay}.


\subsection{Scenarios from higher-dimensional Operators}

We now discuss scenarios with the SM extended by one of the operators listed in Table~\ref{tab:O6R}, possibly 
with more than one index structure. Here, we assume the type-I seesaw relation, so that the effects from the 
active-sterile neutrino mixing are negligible\footnote{That is, here we practically neglect the interactions of the 
minimal model discussed in the previous subsection.}. We consider the following operators in turn: 
$\mathcal O_{Qu\nu L}$, $\mathcal O_{L\nu Qd}$, $\mathcal O_{L\nu Le}$, $\mathcal O_{H\nu e}$ and 
$\mathcal O_{LdQ\nu}$. While we refrain from specifying a UV-complete model, which is not necessary for the 
low-energy phenomenology we are after, it is worthwhile to mention that these operators can easily be obtained  
in leptoquark models ($\mathcal O_{LdQ\nu}$), models with $Z'$ bosons $(\mathcal O_{Qu\nu L}$, $\mathcal 
O_{L\nu Qd}$, $\mathcal O_{L\nu Le}$), and left-right symmetric models ($\mathcal O_{H\nu e}$). We refer to 
Ref.~\cite{DeVries:2020jbs} for a more detailed discussion. 

The above operators are special in the sense that each of them can induce both the production and decay of 
sterile neutrinos by turning on only one flavor configuration. For example, with $\mathcal O_{Qu\nu L}^{11\nu_R3}$, 
we can have the decay chains,
\begin{equation}
\tau^-\to\nu_R+(\pi^-, K^-),\qquad \nu_R\to\nu_\tau+(\pi^0\,, \eta\,, \eta^\prime)\,,
\end{equation}
where the production of the sterile neutrino from $\tau^-$ decays can be associated with a $K^-$, because the 
left-handed down-type quarks $d_L$ are not in their mass eigenstates. In choosing the flavor combinations, we 
focus exclusively on first-generation quarks.

For the operator $\mathcal O_{du\nu e}$, one single flavor setting cannot account for the production and decay 
of the sterile neutrino simultaneously. Hence, we choose $\mathcal O^{11\nu_R3}_{du\nu e}$ for the production 
and $\mathcal O^{11\nu_R1}_{du\nu e}$ for the decay, and assume that $C^{11\nu_R3}_{du\nu e}=C^{11\nu_R1}
_{du\nu e}$. All the possible modes are listed in Table~\ref{tab:modes}.

\begin{figure}[t]
	\centering
	\includegraphics[width=1.1\textwidth]{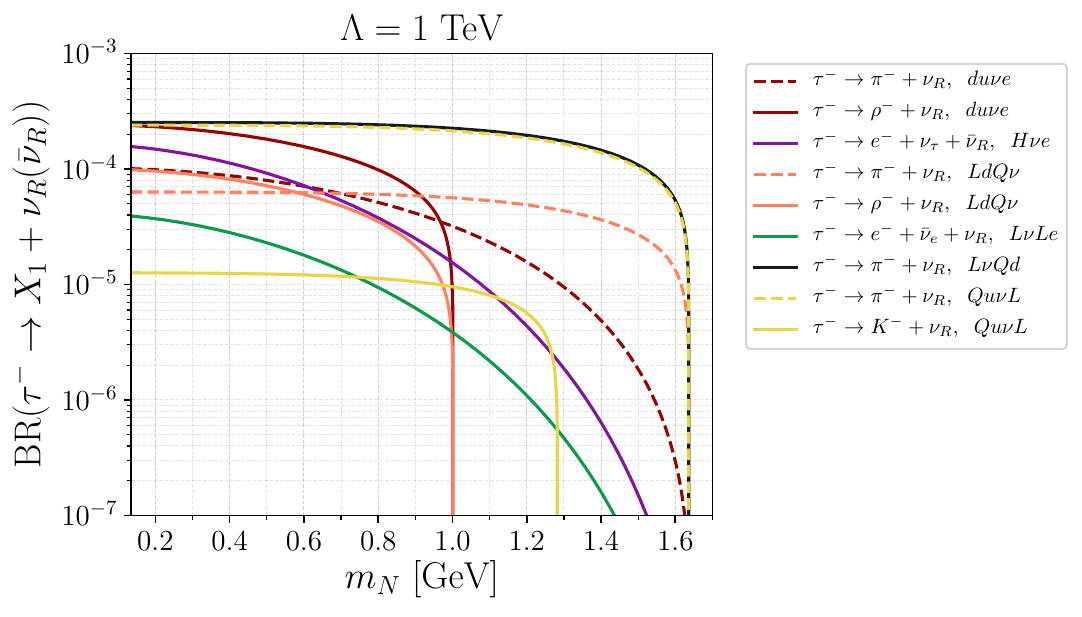}
	\caption{Tau decay branching ratios into a sterile neutrino for $\Lambda = 1$ TeV.} \label{fig:BRtau}
\end{figure}

\begin{figure}[t]
	\centering
	\includegraphics[width=0.85\textwidth]{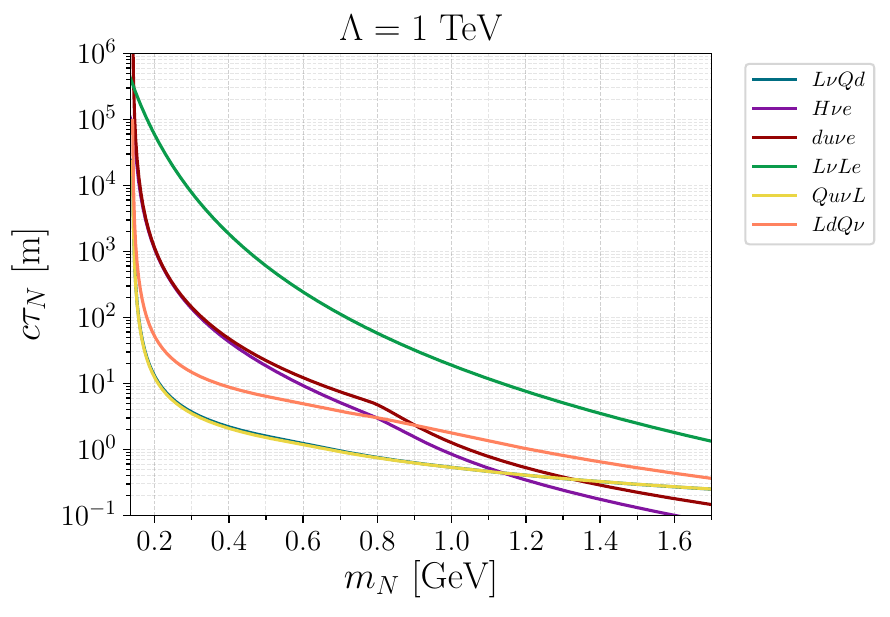}
	\caption{Proper decay lengths of the sterile neutrino in various EFT scenarios for $\Lambda= 1$ TeV.} \label{fig:ctau}
\end{figure}

\begin{figure}[t]
	\centering
	\includegraphics[width=0.48\textwidth]{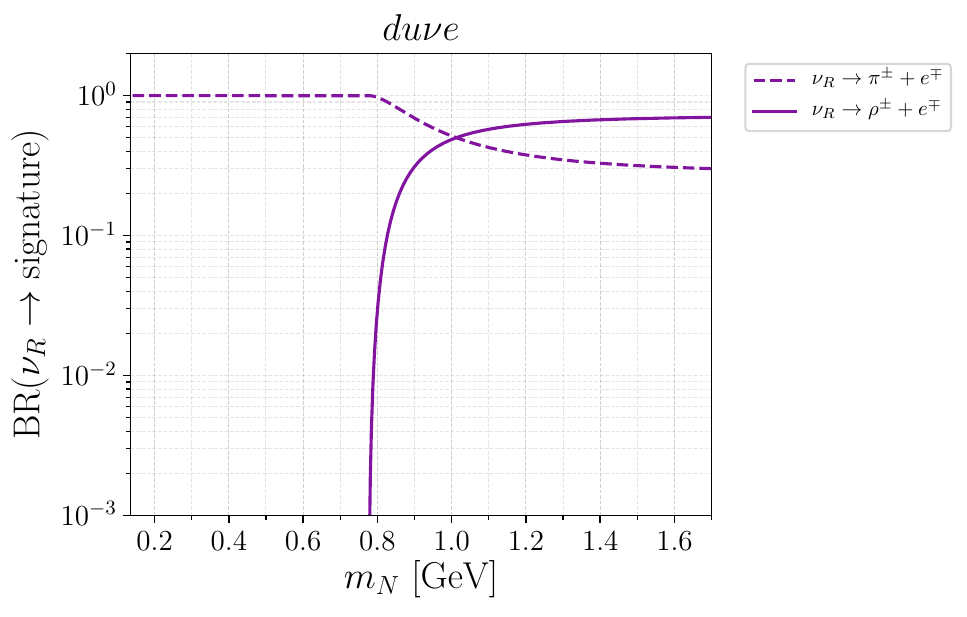}
	\includegraphics[width=0.5\textwidth]{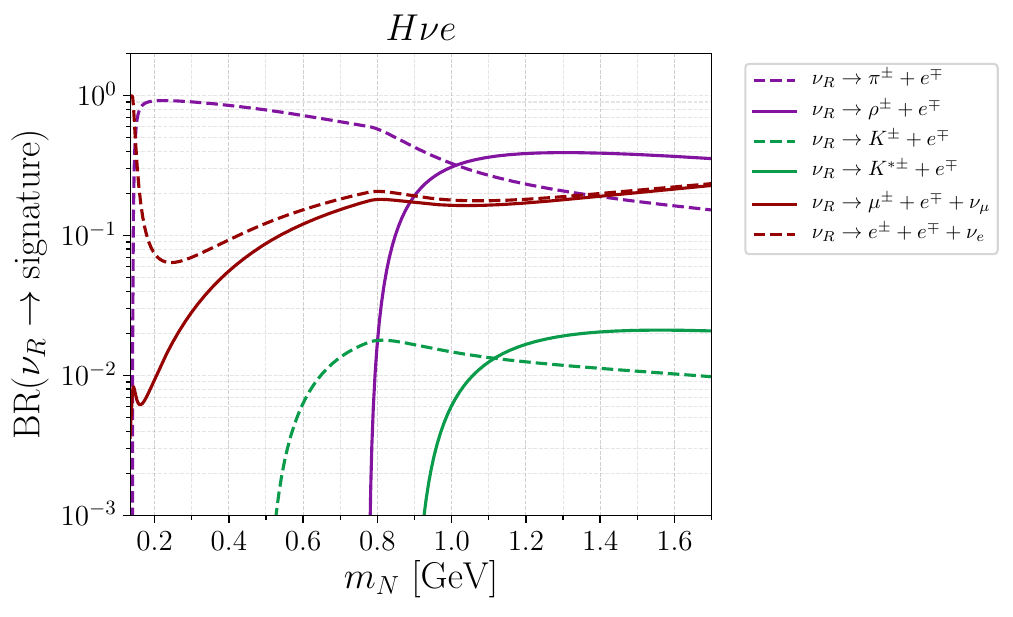}
	\includegraphics[width=0.5\textwidth]{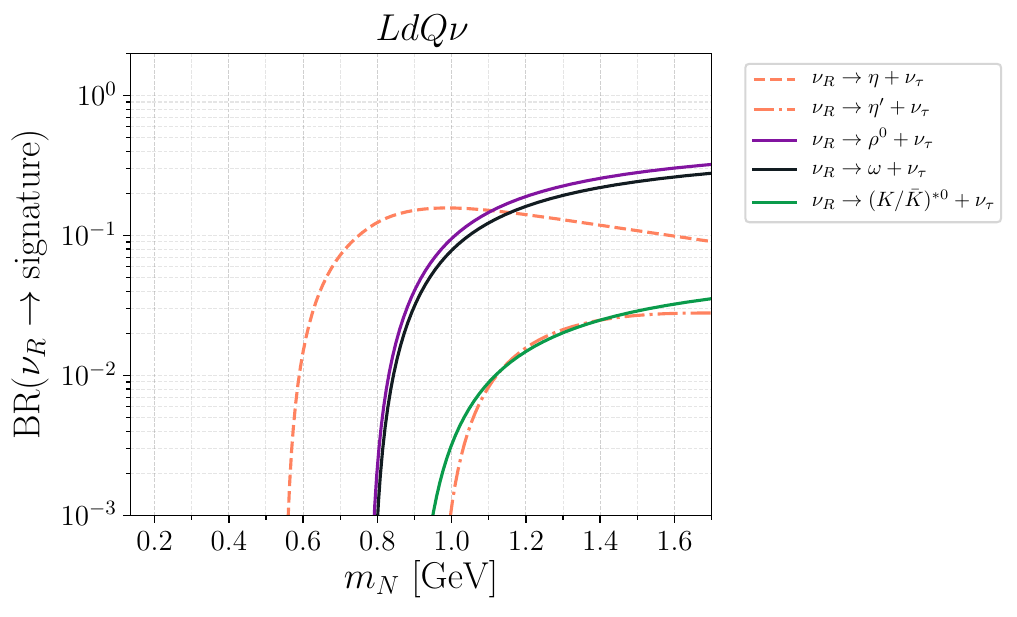}
	\includegraphics[width=0.47\textwidth]{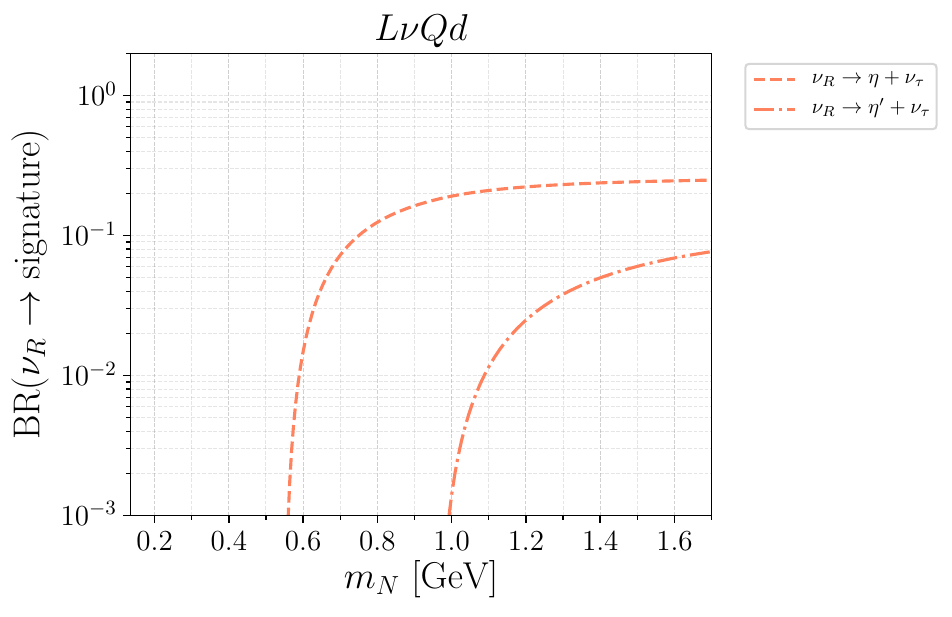}
	\includegraphics[width=0.5\textwidth]{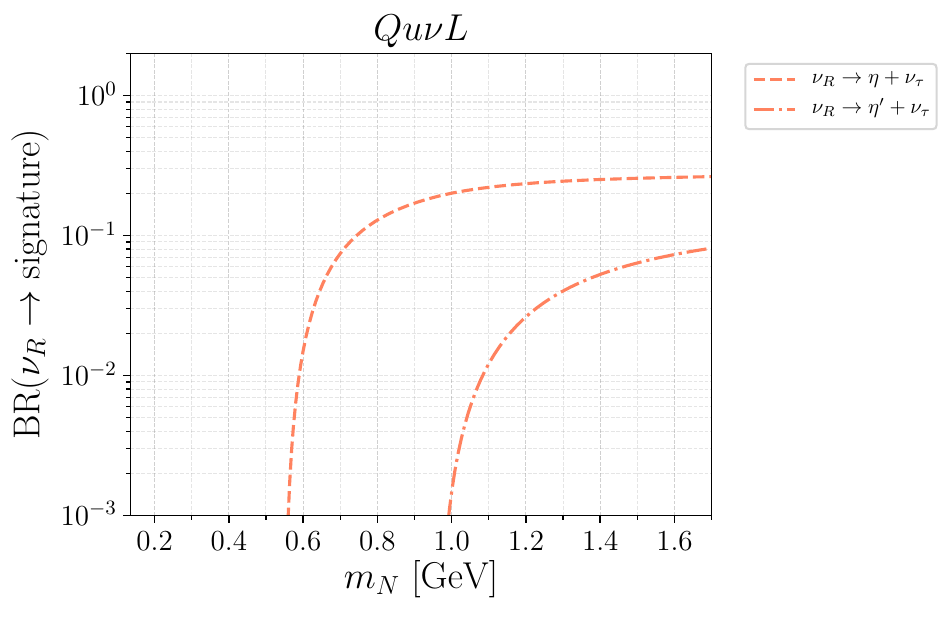}
	\caption{Branching ratios of visible decay modes for the sterile neutrino in $\nu$SMEFT scenarios.} \label{fig:BRN}
\end{figure}

For each dimension-six operator in Table~\ref{tab:O6R}, we suppose their Wilson coefficients are given by $1/\Lambda^2$.
The production and decay rates are thus proportional to $1/\Lambda^4$ and are functions of both $m_N$ and $\Lambda$.
In Fig.~\ref{fig:BRtau} and Fig.~\ref{fig:ctau}, we present respectively the $\tau$ decay branching ratios into a sterile neutrino plus 
anything and the proper decay lengths, $c\tau_N$, of the sterile neutrino, as functions of $m_N$, for a fixed value of $\Lambda=1$ TeV.
All the EFT scenarios summarized in Table~\ref{tab:modes} are included.
Further, we show the branching ratios of the visible decay modes of the sterile neutrino in Fig.~\ref{fig:BRN}.
The branching ratio of the signature decay mode in the scenario with the operator $\mathcal O^{1\nu_R 31}_{L\nu Le}$ is not included here, as it is $100\%$.

We note that in a recent paper~\cite{Beltran:2021hpq}, a phenomenological study on the EFT scenarios with the higher-dimensional operators listed in Table~\ref{tab:modes}, except $\mathcal{O}_{H\nu e}$, has been performed for the LHC with a search strategy based on a displaced vertex, for $m_N\gtrsim 5$ GeV.
In contrast, in this paper we focus on $m_N$ below the $\tau$-lepton mass.

\section{Experiment and Simulation}\label{sec:simulation}

\texttt{Belle~II} is an ongoing experiment at the SuperKEKB accelerator, which is an electron-positron 
collider operated at a relatively low center-of-mass energy $\sqrt{s}=10.58\,$GeV, \textit{i.e.} at the 
$\Upsilon(4s)$ resonance. At \texttt{Belle~II} an electron beam of energy 7\,GeV collides asymmetrically 
with a positron beam of energy 4\,GeV. With a projected 50 ab$^{-1}$ integrated luminosity, besides the 
large number of $B$-mesons, \texttt{Belle~II} is expected to produce inclusively $4.6\times 10^{10}$
tau pairs, via $e^- e^+ \to \tau^-\tau^+$.
These events can be easily tagged, if one of the two $\tau$'s 
decays into one prong. Given the clean environment and the large production rates of the $\tau$'s, \texttt
{Belle~II} provides an ideal avenue for studying rare $\tau$-decays.

For the purpose of this work, we study rare $\tau$ decays into a sterile neutrino, associated with either a 
charged meson or a charged lepton, plus missing energy from an escaping active neutrino.
For the minimal scenario, the sterile neutrino is considered to be mixed with the $\nu_\tau$ only, while for the EFT 
scenarios with higher-dimensional operators the sterile neutrino is assumed to have four-fermion interactions 
with at least one third-generation lepton at the low-energy scale. We focus on the case that the sterile 
neutrinos are long-lived and decay to lighter mesons or leptons, disconnected from the production vertex, 
but within the tracker.
We propose a displaced-vertex (DV) search strategy for these reactions, which requires at least two charged 
final-state particles for the signature.
More concretely, the two displaced tracks can stem either directly from the decay of the long-lived 
sterile neutrino, or from the subsequent \textit{prompt} decay of a meson, such as $\rho^0$ and $\eta$, produced 
from the sterile neutrino's decay.

We now explain the event selections we impose for this search. We define a fiducial volume of the \texttt
{Belle~II} detector by $10\,\text{cm} < r < 80\,\text{cm}$ and $-40\,\text{cm} < z < 120\,\text{cm}$, where $r$ and
$z$ are the transverse and longitudinal distances to the IP, respectively. The positive $z$ direction is defined to 
be on the side of the incoming positron beam. The choice of $r >10\,\text{cm}$ ensures that the background 
events from $K_S$ decays, prompt tracks, as well as detector material interactions are removed. The sterile 
neutrino is required to decay inside the fiducial volume. Second, in general we expect the tracking efficiency to 
deteriorate with increasing (transverse) distance from the IP inside the tracker. To parameterize this effect, we 
apply a naive linear function to interpolate the displaced-tracking efficiency, ranging from 100\% at $r=10\,$cm to 
0\% at $r=80\,$cm; see also Refs.~\cite{Bertholet:2021hjl,Cheung:2021mol}.

The efficiency to reconstruct a DV relies on the final-state tracks. For the case of two tracks stemming from a DV, 
we follow Ref.~\cite{Dey:2020juy} to take this DV reconstruction efficiency to be $12\%$. With any $\pi
^0$, $K^0$, or photon \textit{additional} to the two tracks in the sterile neutrino decay products, the DV reconstruction efficiency is further multiplied 
with $70\%$. Similarly, for any additional pair of charged pions, the efficiency is modified by a factor of $85\%$\footnote{These efficiencies are conservative estimates based on track finding efficiency investigations of the B-factory experiment
 BABAR (see Ref. \cite{Allmendinger:2012ch,BaBar:2013byz}). With the help of various Monte-Carlo event generators the efficiencies of 
reconstructing processes such as $e^+e^-\rightarrow \tau^+\tau^-$, $e^+e^-\rightarrow \pi^+\pi^-(\pi^+\pi^-) \gamma_{ISR}$, where $\gamma_{ISR}$ 
is a high energetic photon emitted from an initial lepton, and hadronic decay modes are evaluated and compared to data of each BABAR run.}
\footnote{To apply the analysis to reduce background events described in the following paragraph, the final state must be reconstructed as detailed as possible. Due to the unobservable neutrino in the final state we can not fully reconstruct it, but every other charged particle should be tracked.}.

On top of all these cuts and efficiencies, we expect that a more detailed analysis including all detector 
effects can remove the remaining SM background events, while retaining about 75\% of the signal events. We
therefore apply an overall efficiency of 75\% on top of the previously mentioned factors.
This estimate is inspired by Ref.~\cite{Dib:2019tuj}, where the authors showed that by computation with the 
four-momenta of the final-state particles it is possible to derive the $\tau$ energy and the LLP mass up to a two-fold 
ambiguity at \texttt{Belle~II}. 
By comparing their distributions it is possible to remove the entire background events 
while keeping $\gtrsim 75\%$ of all the signal events.

The final expected signal-event number can thus be computed as:
\begin{eqnarray}
N_S = 2\cdot N_{\tau\bar{\tau}} \cdot \text{BR}(\tau\to\text{1 prong})\cdot \text{BR}(\tau \to \nu_R+X_1) \cdot \epsilon 
\cdot \text{BR}(\nu_R\to \text{visible})\,,
\end{eqnarray}
where $N_{\tau\bar{\tau}}=4.6\times 10^{10}$, BR$(\tau\to$1 prong$)\approx 85\%$, and $\epsilon$ denotes the 
final event selection efficiency. The factor 2 arises because in each signal event two $\tau$'s are produced, which
can potentially each decay into a sterile neutrino. BR$(\nu_R\to\text{visible})$ is the decay branching ratio 
of the sterile neutrino into at least two charged particles. 
This excludes final states $X_2$ (see Table~\ref{tab:modes}) consisting of neutrinos only, involving 
$\pi^0$, which decays mostly into two photons, or involving $K^0$ (\textit{i.e.} $K_S$ or $K_L$), which is itself also 
long-lived and hence predominantly escapes the detector's fiducial volume.

We perform Monte-Carlo simulations with \texttt{Pythia8.245}, in order to numerically determine the event selection 
efficiencies $\epsilon$ for each benchmark scenario. \texttt{Pythia8} can generate $e^-e^+\to \tau^-\tau^+$ events 
including the effects of ISR (initial state radiation) and FSR (final state radiation). The simulated $\tau$'s are all 
exclusively set to decay to $\nu_R+X_1$, according to the computed branching ratios of BR$(\tau \to \nu_R+X_1)$ 
for different candidates of $X_1$, \textit{cf.}~Table~\ref{tab:modes}. With \texttt{Pythia8} providing the kinematics of 
each simulated sterile neutrino, we estimate its decay probability inside the fiducial volume folded with the linear displaced-tracking efficiency. The DV reconstruction efficiencies which depend on the final state particles are then multiplied with the cutflow efficiency, according to the various sterile neutrino decay branching ratios. At the end, we include the final overall efficiency $75\%$ for removing the background events.

\section{Numerical Results}\label{sec:results}

As discussed in Sec.~\ref{sec:simulation}, with the proposed search strategy, and for 50 ab$^{-1}$ integrated 
luminosity, we expect vanishing background at \texttt{Belle~II}. In our numerical results, we show three-signal-event isocurves as the exclusion limits at 95\% confidence level, and also as a measure of sensitivity of the experiment to the $\nu_R$-models.

\begin{figure}[t]
	\centering
	\includegraphics[width=0.55\textwidth]{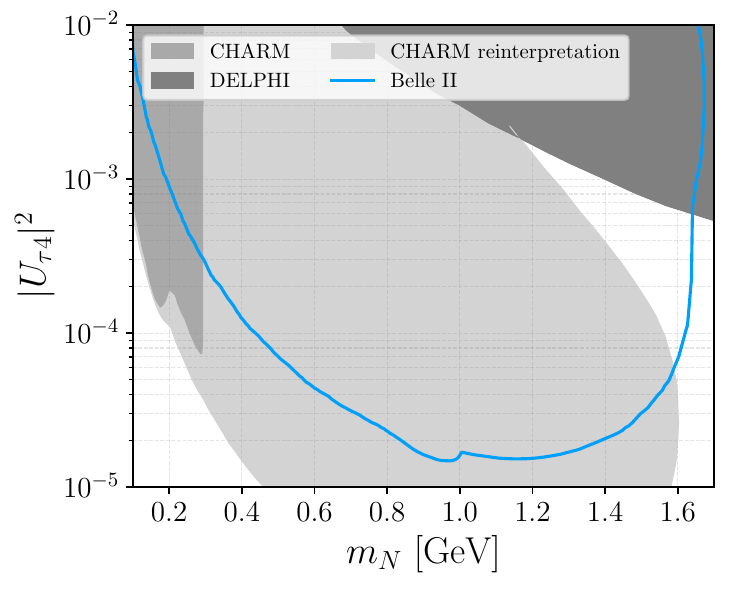}
	\caption{Expected sensitivity limits on the mixing matrix element squared, $|U_{\tau 4}|^2$, as 
	function of the sterile neutrino mass, for the minimal scenario.
	  The dark and medium gray areas correspond to the parameter regions currently excluded by   \texttt{DELPHI}~\cite{DELPHI:1996qcc} and \texttt{CHARM}~\cite{CHARM:1985nku,Orloff:2002de}, respectively. A recent re-interpretation~\cite{Boiarska:2021yho} of the \texttt{CHARM} experiment results~\cite{CHARM:1983ayi,CHARM:1985nku} further excludes the light gray parameter region.
	The kink at $m_N\sim1\,$GeV is due to the $\rho$-threshold in the $\tau$-decay.}\label{fig:sensitivity_minimal}
\end{figure}

In Fig.~\ref{fig:sensitivity_minimal}, we present the sensitivity limits for the minimal scenario, \textit{cf.}
Sec.~\ref{subsec:minimal}, shown in the plane $|U_{\tau 4}|^2$ vs.~$m_N$. We find general agreement with the 
exclusion limits obtained in Ref.~\cite{Dib:2019tuj}. We also compare our limits with existing bounds
obtained by the \texttt{DELPHI}~\cite{DELPHI:1996qcc} (marked dark gray) and  \texttt{CHARM}~\cite{CHARM:1985nku,Orloff:2002de} (marked medium gray) experiments, respectively.
Moreover, recently Ref.~\cite{Boiarska:2021yho} has performed a re-analysis of the CHARM search results~\cite{CHARM:1983ayi,CHARM:1985nku}, obtaining updated bounds on $|U_{\tau 4}|^2$ in the  sterile neutrino mas range 290 MeV $< m_N < 1.6$ GeV.
	We have included these exclusion limits in Fig.~\ref{fig:sensitivity_minimal}, shown in light gray. 
	We find most of the parameter space that \texttt{Belle~II} is sensitive to has now been excluded, except a relatively limited region at $1.2\text{ GeV}\lesssim m_N \lesssim 1.7$ GeV for $|U_{\tau 4}|^2 \sim \mathcal{O}(10^{-4})$.
For values of $|U_{\tau 4}|^2$ smaller than the \texttt{Belle~II} limits, the production rates of the sterile neutrinos become too small and the sterile neutrinos are too long-lived to decay inside the detector fiducial volume, resulting in fewer than three signal events predicted.
 The left and right ends of the exclusion limits are determined by kinematical thresholds.
 At $m_N\sim 1.0\,$GeV the isocurve displays a kink. 
This is due to the threshold of a $\tau$-decay mode into a 
sterile neutrino and a rho-meson.

\begin{figure}[t]
	\centering
	\includegraphics[width=0.49\textwidth]{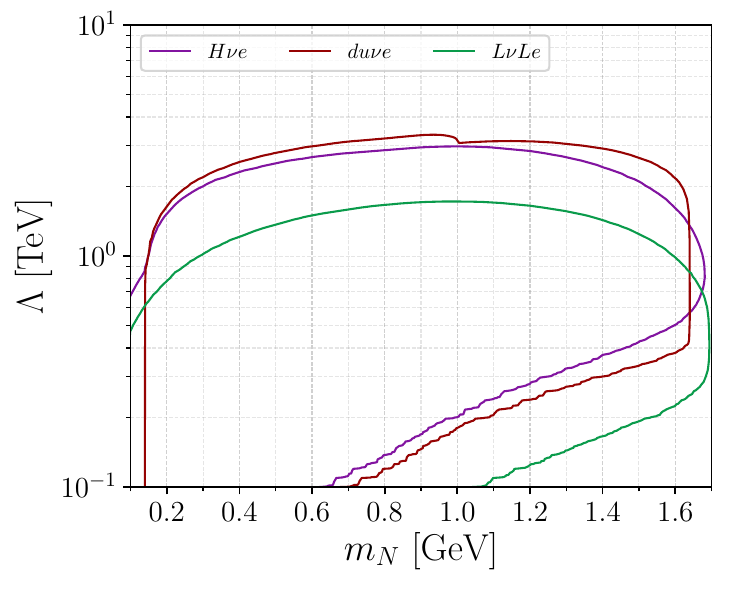}
	\includegraphics[width=0.49\textwidth]{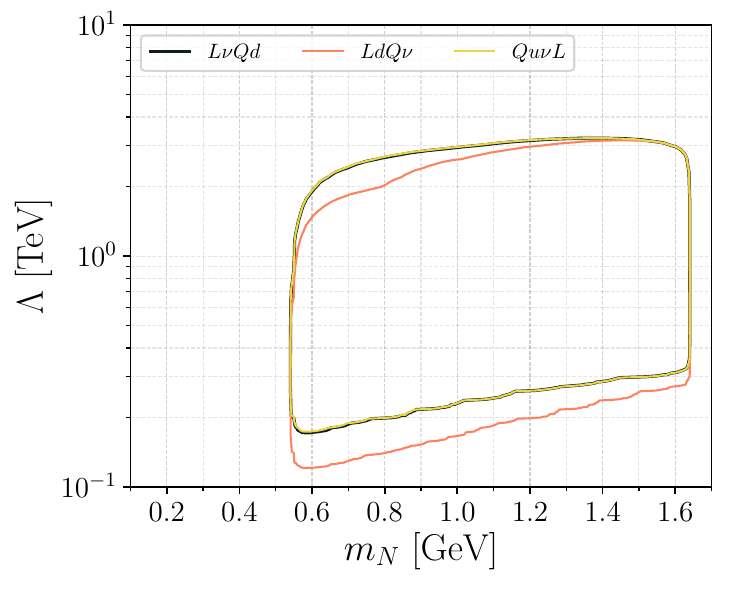}
	\caption{Sensitivity limits for the various EFT scenarios.
			We fix the Wilson coefficients at 1 and show projected bounds on the new-physics scale, 
			$\Lambda$, as functions of $m_N$.} \label{fig:sensitivity_eft}
\end{figure}

The \texttt{Belle~II} exclusion limits for the various EFT operators listed in Sec.~\ref{sec:model} are presented in 
Fig.~\ref{fig:sensitivity_eft}, in the plane $\Lambda$ vs.~$m_N$. The left plot collects results for operators that are 
sensitive to neutrino masses below $\sim m_\eta$ as they induce sterile neutrino decays into pions or charged leptons.
The right plot displays operators that are insensitive to sterile neutrino masses below $\sim m_\eta$. We find the $C_
{L\nu Q d}$ (black) and $C_{Qu\nu L}$ (yellow) sensitivities are almost identical, because the production and decay 
rates of the sterile neutrinos are similar (\textit{cf.}~Eqs.~\eqref{d6NC}-\eqref{match6NC} and Eq.~\eqref{PSM}). In general, we find all the six considered operators can be probed up to $\sim 1-3\,$TeV in $\Lambda$ across the sensitive mass ranges in the long-lived regime (large $\Lambda$).
For even larger $\Lambda$ values, the sterile neutrino lifetime would become so long that they decay only after traversing the detector and their production rates are also reduced too much, while for $\Lambda\lesssim100\,$GeV they decay before reaching the fiducial volume. For such small values of $\Lambda$ the $\nu$SMEFT framework is inapplicable. 
For the same reason as in the minimal scenario (see Fig.~\ref{fig:sensitivity_minimal}) we observe a kink at $m_N$ 
about 1.0\,GeV in the long-lived regime for the $\mathcal{O}_{d u \nu e}$ and $\mathcal{O}_{Ld Q \nu}$ operators.  For $LdQ \nu$ the kink is not as pronounced as in $du\nu e$ or the minimal 
scenario.
Comparing the $\tau\rightarrow \nu_R + (\pi,\rho)$ 
branching ratios (see Fig.~\ref{fig:BRtau}) for the EFT operators, we see that the rho branching ratio of $du\nu e$ 
quickly dominates the $m_N \simle 1\,$GeV region over the pion branching ratio, whereas for $LdQ \nu$ the rho only 
overtakes for $m_N\simle  0.6\,$GeV. Thus, the kink is less pronounced in this scenario.

\begin{figure}[t]
	\centering
	\includegraphics[width=0.5\textwidth]{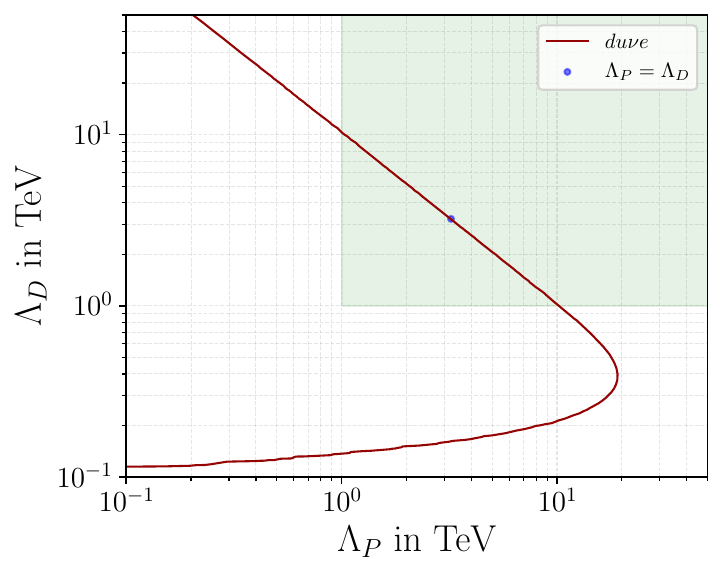}
	\caption{Sensitivity limits for the $du\nu e$ scenario for a fixed neutrino mass $m_N=\SI{0.8}{GeV}$. 
	The respective Wilson coefficients are fixed to 1, allowing to show projected bounds on the new physics 
	scales $\Lambda_P$ and $\Lambda_D$ responsible for production and decay of the neutrino, respectively. Further included are 
	a green area marking the parameter region $\Lambda_{P/D}\geq \SI{1}{TeV}$, for which the $\nu$SMEFT framework 
	is applicable, and a blue dot depicting the limit for $\Lambda_P=\Lambda_D$ corresponding to the limit of 
	Fig.~\ref{fig:sensitivity_eft} at $\SI{0.8}{GeV}$.}\label{fig:sensitivity_individual_couplings}
\end{figure}

In general UV-complete scenarios, several EFT operators can be induced simultaneously. In such scenarios the limits could change somewhat from our results here. While a general scan is not very useful, we have estimated what happens in a specific scenario where we have independent couplings for the production and decay of the sterile neutrino. In Fig. \ref{fig:sensitivity_eft} we show constraints as a function of $m_N$ for the choice $C_{duve}^{11\nu_R3} =C_{duve}^{11\nu_R1}$, where the first (second) coupling dominates the production (decay) of the sterile neutrino. We label the corresponding $\Lambda's$ by $\Lambda_P$ (for production) and $\Lambda_D$ (for decay).  For a fixed sterile neutrino mass of $m_N=\SI{0.8}{GeV}$, we investigated what happened if we freely vary the two couplings. We observe that in the window where $\Lambda_{P,D} \geq 1$ TeV (where the EFT is valid) the constraints on the individual couplings can be strengthened by roughly a factor of 3, at the cost of a reduced sensitivity to the other coupling. With the expected sensitivities we therefore do not expect a significant difference in sensitivity for scenarios with more EFT couplings. Of course, specific UV-complete models can be studied in detail by matching to our EFT operators.

To the best of our knowledge, no other constraints on these EFT couplings have been established for sterile neutrino masses in this region.
For sterile neutrino masses below $100\,$MeV, several operators considered in this work have been constrained from LHC searches for tau production plus missing transverse energy \cite{Alcaide:2019pnf}.
The resulting constraints are at the TeV level as well, but, as mentioned above, require lighter sterile neutrinos in order for them to traverse the LHC detectors.
In addition, constraints from meson decays, tau decays, lepton flavor universality, CKM unitarity, $\beta$-decays, and EC$\nu$NS, have been discussed in Refs.~\cite{Alcaide:2019pnf,Biekotter:2020tbd,Li:2020lba,Li:2020wxi,Mandal:2020htr,Akimov:2017ade,Han:2020pff,Bischer:2019ttk}, where either a massless sterile neutrino was assumed, or $\nu$SMEFT operators with flavor configurations different from those we have studied here were considered.
Consequently, these constraints are not shown in Fig.~\ref{fig:sensitivity_eft}.

\section{Conclusions}\label{sec:conclusions}

At Belle~II, $10^{10}-10^{11}$ tau leptons are predicted to be produced with an integrated luminosity of 50 ab$^{-1}$ over the whole experiment lifetime, making it possible to search for rare $\tau$ decays.
In this work, we have proposed a strategy based on displaced vertex at Belle~II, to search for long-lived sterile neutrinos produced from $\tau$ decays.
The search includes a requirement on the fiducial volume consisting of the tracker and a linear displaced-tracking efficiency.
Further, to reconstruct the displaced vertices, we apply realistic efficiency factors depending on the final states of the sterile neutrino decays.
Finally, based on existing literature~\cite{Dib:2019tuj,Dey:2020juy}, an overall factor of $75\%$ is imposed to account for removing the remaining background events.

We have not only considered the minimal scenario where the sterile neutrinos are produced and decaying via the same mixing parameter, $|U_{\tau 4}|^2$, but also worked in the framework of the Standard Model Effective Field Theory extended with sterile neutrinos encoding the effects of heavy new physics into non-renormalizable operators with dimensions up to six.
Following the proposed search strategy, we obtained the sensitivity limits of Belle~II to these theoretical scenarios.
In the minimal scenario, our results are in general agreement with those obtained in Ref.~\cite{Dib:2019tuj}.
For the EFT scenarios, we switch on one EFT operator at a time, and assume the type-I seesaw relation allowing us to disregard the weak interactions with the mixing parameter which is too small to affect the phenomenology.
We find that with our search strategy for the various $\nu$SMEFT operators considered, Belle~II can probe the new-physics scale up to about 3 TeV, assuming unity Wilson coefficients, in the kinematically allowed mass range, proving Belle~II has unique sensitivities to $\nu$SMEFT interactions with third-generation leptons.

\bigskip
\section*{Acknowledgements}
\bigskip
We thank Florian Bernlochner, Claudio Dib, Juan Carlos Helo, Nicol\'as Neill, Maksym Ovchynnikov, Abner Soffer, and Arsenii Titov for useful discussions.
Z.S.W. is supported by the Ministry of Science and Technology (MoST) of Taiwan with grant numbers MoST-109-2811-M-007-509 and MoST-110-2811-M-007-542-MY3. Financial support for H.K.D. by  the DFG (CRC 110, ``Symmetries and the Emergence of Structure in QCD'') is gratefully 
acknowledged.

\bigskip

\appendix
\section{Two-body Decay Processes with a Sterile Neutrino}\label{appendix:twobodydecay}
\subsection{Charged Currents}
Via the charged-current interactions introduced in Eq.~\eqref{d6CC}, the $\tau$ lepton and sterile neutrino can undergo two-body 
decays into final states consisting of a lepton and a charged pseudoscalar or vector meson.  For a pseudoscalar meson $M_{ij}$ 
consisting of the valence quarks, $\bar{q}_i$ and $q_j$, we define the matrix element for the axial-vector current as
\be
\langle 0|\bar{q}_i\gamma^\mu \gamma^5 q_j|M_{ij}(q)\rangle \equiv iq^\mu f_{M_{ij}}\,,
\ee
where $q$ is the momentum of $M_{ij}$ and $f_{M_{ij}}$ is the decay constant.
After applying the equation of motion to the current, we define the decay constant for the axial current,
\begin{equation}\label{PSM}
\langle0|\bar{q}_i \gamma^5 q_j|M_{ij}(q)\rangle =i\frac{m^2_{M_{ij}}}{m_{q_i}+m_{q_j}}f_{M_{ij}}\equiv if^S_{M_{ij}}\,,
\end{equation}
where $m_{q_i}$ and $m_{M_{ij}}$ are the masses of quark $q_i$ and pseudoscalar meson $M_{ij}$, respectively.
The matrix elements of the vector and tensor currents with a vector meson are given as:
\begin{equation}
\begin{aligned}
\langle0|\bar{q}_i\gamma^\mu q_j|M_{ij}^*(q,\epsilon)\rangle&\equiv i f^V_{M_{ij}} m_{M_{ij}^{*}}\epsilon^\mu\,,\\
\langle0|\bar{q}_i\sigma^{\mu\nu} q_j|M_{ij}^*(q,\epsilon)\rangle&\equiv -f^T_{M_{ij}}(q^\mu\epsilon^\nu-q^\nu \epsilon^\mu)\,,\\
\end{aligned}
\end{equation}
where $M_{ij}^*(q,\epsilon)$ denotes a vector meson with mass $m_{M^*_{ij}}$, momentum $q$, and polarization vector $\epsilon$, and we assume $ f^T_{M_{ij}} \simeq f^V_{M_{ij}} $.
We list the values of all the relevant decay constants in Table~\ref{table:ConstantC}.
 \begin{table}[t]
	\centering
	\begin{tabular}{||c | c | c |c ||  }
		\hline
		meson $M_P$ &  $f_M$ [MeV] & meson $M_V$ &  $f^V_M$ [MeV]\\
		\hline 
		
		$K^\pm$         & 155.6 \cite{Rosner:2015wva}		& $K^{*\pm}$     & 230 \cite{Dreiner:2006gu}	\\
		\hline
		$\pi^\pm$	     &  130.2 \cite{Rosner:2015wva}	& $\rho^\pm$   &209 \cite{Ebert:2006hj}	\\	
		\hline 		
	\end{tabular}
	\caption{Decay constants for charged pseudoscalar and vector mesons.}
	\label{table:ConstantC}
\end{table}

\subsection{Neutral Currents}
\subsubsection{The Two-body Decay of the Sterile Neutrino in the Minimal Model}

In the minimal model, the sterile neutrino can decay into a neutral pseudoscalar meson $M^0_P$ or vector meson $M^0_V$ 
via the mixing with $\nu_\tau$. The decay width of $\nu_R\rightarrow \nu_\tau +M^0_P$ can be written as~\cite{DeVries:2020jbs}
\begin{equation}
\Gamma (\nu_R\rightarrow \nu_\tau M^0_{a,P})= 2\times \frac{G^2_F f_a^2 m_N^3 |U_{\tau4}|^2}{32 \pi} \left(1-\frac{m_{a}^2}{m_N^2}\right)^2\,,
\end{equation}
where $a$ denotes $\pi^0, \eta,$ or $\eta^\prime$, $m_a$ is the mass of meson $M^0_{a,P}$, and we include a factor 2 explicitly to account for the charge-conjugated decay modes of the Majorana sterile neutrino (similarly for the other decay rates expressions given below).
For $\nu_R \rightarrow \nu_\tau +M^0_V$, the decay rates are~\cite{DeVries:2020jbs}
 \begin{equation}
 \Gamma(\nu_R\rightarrow \nu_\tau M^0_{a,V})=2 \times \frac{G_F^2  f_a^2 g_a^2 |U_{\tau 4}|^2 m_N^3}{32\pi }
 \left(1+2 \frac{m_a^2}{m_N^2}\right)\left(1-\frac{m_a^2}{m_N^2}\right)^2\,,
 \end{equation}
 where $a=\rho^0, \omega$, or $\phi$, and $m_a$ labels the mass of meson $M^0_{a,V}$.  In Table~\ref{table:ConstantN}, we list 
 the values of $f_a$ and $g_a$ we use for the numerical studies in this work. 
   
 \begin{table}[t]
 	\centering
 	\begin{tabular}{||c | c | c||}
 		\hline
 		meson $M^0$ & $f_a$ [Me$\text{V}$] & $g_a$ \\
 		\hline 
 		$\rho^0$     &  220.6 \cite{Coloma:2020lgy} &  $1-2\sin^2\theta_w$ \\
 		$\omega$	  & 198 \cite{Coloma:2020lgy}  & $-\frac{2}{3}\sin^2\theta_w$ \\
 		$\phi$	  & 227.4 \cite{Coloma:2020lgy}  & $\sqrt{2}(-\frac{1}{2}+\frac{2}{3}\sin^2\theta_w)$ \\ 
 		\hline
 		$\eta$ & 81.7 \cite{Escribano:2015yup}&
 		
 		\\
 		\cline{1-2}
 		$\eta^\prime$ & -94.7 \cite{Escribano:2015yup}& 		
Not applicable 		\\
 			\cline{1-2}
 		$\pi^0$ & 130 \cite{Coloma:2020lgy}&
 		
 		\\
 		\hline
 	\end{tabular}
 	\caption{Decay constants and $g_a$ of neutral  mesons.  }
 	\label{table:ConstantN}
 \end{table}

\subsubsection{The Two-body Decay of the Sterile Neutrino via higher-dimensional Operators}
Some neutral-current interactions listed in Eq.~\eqref{d6NC} include a sterile neutrino and two quarks, and can hence induce decays of the 
sterile neutrino into a neutral pseudoscalar meson. We reproduce them here:  
\begin{equation}
\mathcal L_{NC}= \frac{2G_F}{\sqrt{2}}\big(\bar{u}_R u_L \bar{\nu}_L c^{NC}_{\rm SLR} \nu_R +\bar{d}_L d_R \bar{\nu}_L c^{NC1}_{\rm SRR} \nu_R \big)+ \rm h.c.
\end{equation} 
To compute the decay rates of the sterile neutrino via these terms, we work in the $SU(3)$ chiral perturbation theory, following the calculation 
procedure as detailed in Ref.~\cite{Gu:2018swy}. We first write down the leading-order chiral Lagrangian containing the Lorentz- and 
chiral-invariant terms with the lowest number of derivatives,
\be\label{eq:chiral}
\mathcal L_{\pi, K} = \frac{F^2}{4 } \mathrm{Tr}\left[(D_\mu U)^\dagger (D^\mu U)\right]+ \frac{F^2}{4} \mathrm{Tr}\left[U^\dagger \chi + U \chi^\dagger\right]\,,
\ee
where $D_\mu U = \partial_\mu U - i l_\mu U + i U r_\mu\,$ and  
$\chi = 2 B (M + s -ip)\,$.  $l_\mu, r_\mu, s, p$ are external sources  and $M =\mathrm{diag} (m_u, m_d,m_s)$ is a diagonal $3\times 3$ quark mass matrix. $U$ is given by
\be
U(x) = \mathrm{exp}\left(\frac{i\sqrt{2}\Pi(x)}{\ F}\right)\,,\qquad \Pi(x) = \bma \frac{\pi^0}{\sqrt{2}}+\frac{\eta_8}{\sqrt{6}}+\frac{\eta_0}{\sqrt{3}} &\pi^+ & K^+\\\pi^-&-\frac{\pi^0}{\sqrt{2}}+\frac{\eta_8}{\sqrt{6}}+\frac{\eta_0}{\sqrt{3}} & K^0 \\
K^- & \bar{K}^0 &-\sqrt{\frac{2}{3}}\eta_8+\frac{\eta_0}{\sqrt{3}} \ema .
\end{equation}
$\eta_8$ and $\eta_0$ are in the singlet-octet basis and their relations with the physical states $\eta$ and $\eta^\prime$ are
\begin{equation}\label{mixing}
\bma \eta\\ \eta^\prime \ema =  \frac{1}{F}\bma F_8 \cos \theta_8 & -F_0 \sin \theta_0\\ F_8 \sin \theta_8 & F_0 \cos \theta_0\ema \bma \eta_8\\\eta_0\ema\,.
\end{equation}
The values of the relevant parameters are~\cite{Gu:2018swy,Chen:2014yta}
\bea
F&=&92.2 \,\text{MeV}\,, \qquad F_0=118.1 \,\text{MeV}\,,\qquad F_8 =133.8 \,\text{MeV}\,,\nn\\
\theta_0 &=&-11.0\si{\degree}\,,\, \qquad\qquad\theta_8= -26.7\si{ \degree}\,.
\eea
By using the external source field method, we find
\bea
s+ip&=&\frac{-2G_F}{\sqrt{2}}\Big\{\bar{\nu}_\tau c^{NC}_{\rm SLR} \nu_R+ (\bar{\nu}_\tau c^{NC1}_{\rm SRR} \nu_R)^\dagger \Big\}\,,\nn\\
s-ip&=&(s+ip)^\dagger\,.
\eea
To obtain the decay rates of the sterile neutrino,  we insert these currents into Eq.~\eqref{eq:chiral} and expand $U$ to leading order.

The neutral tensor current in Eq.~\eqref{d6NC} leads to the decay of the sterile neutrino into a vector meson.
We define the following matrix element:
\begin{equation}
\langle0|\bar{q}_1\sigma^{\mu\nu} d|M_{V}^0(q,\epsilon)\rangle\equiv -f^T_{M_{}}(q^\mu\epsilon^\nu-q^\nu \epsilon^\mu)\,,
\end{equation}
where $q_1$ can be a down or strange quark, $d$ is a down quark, and $M^0_{V}=\rho^0\,, \omega\,, K^{*0}$. The tensor decay constants 
$f^T_{M}$ are related to the vector decay constants by $f^T_{\rho^0} = f_{\rho^\pm} /\sqrt{2}$,  $f^T_{\omega} = f_{\omega} /\sqrt{2}$, and $f^T
_{K^{*0}}=f_{K^{*\pm}}$.

\section{Three-body Decays}\label{appendix:threebodydecay}
In the minimal scenario, the sterile neutrino can decay into three light active neutrinos, and the corresponding decay rates can be expressed 
in a closed form~\cite{Bondarenko:2018ptm}
\begin{equation}
\Gamma (\nu_R \rightarrow\nu_\tau \nu_\beta\bar{\nu}_\beta) =2\times (1+\delta_{\tau \beta}) \frac{G_F^2 m^5_N |U_{\tau 4}|^2}{768\pi^3}\,,
\end{equation}
where $\beta=e, \mu, \tau$ is the flavor of light neutrinos.
However, in most cases the three-body decay widths cannot be computed analytically.
Thus, we compute these three-body phase space integrals numerically.
With the help of \texttt{FeynCalc}~\cite{Shtabovenko:2020gxv,Shtabovenko:2016sxi,Mertig:1990an} and the method explained in Appendix~B of Ref.~\cite{DeVries:2020jbs}, we automatize the calculation procedure in \texttt{Mathematica}.

\bibliographystyle{JHEP}
\bibliography{bibliography}
\end{document}